\documentstyle[prl,aps,multicol,epsf,rotate]{revtex}
\begin{document}
\draft
\preprint{XXXX}
\title{ Numerical Solution of the Mode-Coupling Equations for the 
 Kardar-Parisi-Zhang Equation in One Dimension}
\author{Francesca Colaiori and M. A. Moore}
\address{Department of Physics and Astronomy, University of 
Manchester,
Manchester, M13 9PL, United Kingdom}
\date{\today}
\maketitle
\begin{abstract}
We have studied the Kardar-Parisi-Zhang equation in the strong coupling 
regime in the mode-coupling approximation. 
We solved  numerically in dimension $d=1$ for the correlation function 
at wavevector ${\bf k}$. 
At large times $t$ we found the predicted stretched exponential decay
consistent with our previous saddle point analysis in  [Phys. Rev. E {\bf 63},
057103 (2001)],
but  we also observed that the decay to zero occurred in an unexpected
oscillatory way.   We have compared the  
results from mode-coupling for the scaling functions with the recent 
exact results from Pr\"ahofer and Spohn [cond-mat/0101200] for $d=1$ who
also find an oscillatory decay to zero.
\end{abstract}
\begin{multicols}{2}
%
The Kardar-Parisi-Zhang (KPZ) \cite{KPZ} equation is one of the 
most important phenomenological equations in physics. It is essentially a
non-linear Langevin equation and was  proposed in 1986 as a coarse grained
description of a growing interface. It is the simplest generalization
of the diffusion equation which includes a relevant non-linear 
term, and probably as a consequence of this 
the KPZ equation also arises in  connection with many other
important physical problems (the Burgers equation for 
one-dimensional turbulence \cite{For},
directed polymers in a random medium \cite{Kar,Der,Par} etc.). 

The KPZ equation in the context of a growing interface describes it by a single
valued height function $h({\bf x},t)$ on a $d$-dimensional 
substrate
${\bf x}\in \Re^d$ is:
\begin{equation}
\partial_t h({\bf x},t)=
\nu \nabla^2 h +\frac{\lambda}{2}(\nabla h)^2+\eta({\bf x},t) \,.
\label{KPZ}
\end{equation}
The first term on the right of Eq. (1) represents the  forces which tend to
smooth the interface, the second describes the non-linear growth 
locally
normal to the surface, and the last is a noise term which mimics the
stochastic nature of the growth process \cite{rev}, usually chosen to be
Gaussian, with zero mean and second 
moment
$\langle \eta({\bf x},t)\eta({\bf x'},t')\rangle=
2D\delta^{d}({\bf x}-{\bf x'})\delta(t-t') .$
The equal time  interface profile is usually described in terms
of the roughness:
$w=\sqrt{\langle h^2({\bf x},t)\rangle- \langle h({\bf x},t)\rangle^2} $
which for a system of size $L$ behaves like $L^{\chi} f(t/L^z)$,
where $f(x)\rightarrow const$ as $x\rightarrow \infty$
and $f(x) \sim x^{\chi/z}$ as $x \rightarrow 0$, so that $w$
grows with time like $t^{\chi/z}$ until it saturates to $L^{\chi}$
when $t\sim L^z$. $\chi$ and $z$ are the roughness and dynamic
exponent respectively. 

Above two  dimensions, there are two distinct types of solution of the 
KPZ equation.
In the weak coupling regime ($\lambda < \lambda_c$) the non-linear
term is irrelevant and the behavior
is governed by the Gaussian ($\lambda=0$) fixed point and $z=2$. 
The strong coupling regime ($\lambda > 
\lambda_c$),
where the non-linearity is relevant (and $\lambda_c=0$ for all $d\leq 2$)
is characterised by 
exponents which are not known exactly in general dimension $d$.
>From the Galilean invariance \cite{For}
(invariance of Eq. (\ref{KPZ}) under an infinitesimal
tilting of the surface) one can derive the relation
$\chi+z=2$, which leaves just one independent exponent. For the
special case when $d=1$, the existence of a 
fluctuation-dissipation
theorem gives the exact results 
$\chi=1/2$, $z=3/2$. 

Almost by definition there is no small parameter for a systematic
perturbative treatment of the strong coupling regime. One is forced either
into numerical studies of the KPZ equation, which are naturally difficult
for dimension $d$ greater than two, or into ``ad hoc'' approximations. The
best known of these is the so-called mode-coupling 
approximation \cite{new,Do,usprl}, 
in which in the diagrammatic expansion for the correlation
and response function only diagrams which do not renormalize
the three point vertex $\lambda$ are retained. One of the purposes of this
paper is to investigate the accuracy of the mode-coupling approximation by
comparing it with the recently obtained exact solution of Pr\"ahofer and Spohn
\cite{PS} for $d=1$. The mode-coupling approximation  is gratifyingly
close to the exact solution, which encourages one to believe in the
utility of the mode-coupling approach in higher dimensions where no 
exact solution is known or likely to be found. 

In a recent paper we studied the long time properties of the KPZ 
equation within the mode-coupling approximation,
and we predicted  a stretched exponential decay for
the correlation function at long times.
In this paper we have found numerically the solution of the 
mode-coupling equations in $d=1$ 
which confirms the results of the previous  asymptotic analysis but which
also reveals that the 
correlation functions decay to zero in an oscillatory manner
-- a fact which was not revealed by our previous asymptotic analysis.
 
Mode-coupling equations are coupled equations for
the correlations and response function.
The correlation and response function are defined in
Fourier space by
\begin{eqnarray}
&C({\bf k},\omega)=\langle h({\bf k},\omega) h^{*}({\bf k},\omega) 
\rangle,
\nonumber
\\
&G({\bf k},\omega)=
\langle
\partial h({\bf k},\omega) /\partial \eta({\bf 
k},\omega)
\rangle ,
\nonumber
\end{eqnarray}
\noindent
where $\langle \cdot \rangle$ indicate an average over $\eta$.
In the mode-coupling approximation, the correlation and response
functions are the solutions of two coupled equations,
\end{multicols}
\begin{eqnarray}
&
G^{-1}({\bf k},\omega)=G^{-1}_0({\bf k},\omega)+\lambda^2
\int \frac{d\Omega}{2 \pi} \int \frac{d^dq}{(2 \pi)^d}
\left[{\bf q} \cdot ({\bf k}-{\bf q})\right]\left[{\bf q} \cdot 
{\bf
k}\right]
G({\bf k}-{\bf q},\omega - \Omega) C({\bf q},\Omega) \,,
\label{mc1}
\\&
C({\bf k},\omega)=C_0({\bf k},\omega)+\frac{\lambda^2}{2}
\mid G({\bf k},\omega)\mid^2
\int \frac{d\Omega}{2 \pi} \int \frac{d^dq}{(2 \pi)^d}
\left[{\bf q} \cdot ({\bf k}-{\bf q})\right]^2
C({\bf k}-{\bf q},\omega - \Omega) C({\bf q},\Omega) \,,
\label{mc2}
\end{eqnarray}
\begin{multicols}{2}
\noindent
where $G_0({\bf k},\omega)=(\nu k^2 - i \omega)^{-1}$ is the bare
response function, and $C_0({\bf k},\omega)=2 D \mid G({\bf
k},\omega)\mid^2$.
In the strong coupling limit,
$G({\bf k},\omega)$ and $C({\bf k},\omega)$ take the following
scaling forms
\begin{eqnarray}
&
G({\bf k},\omega)=k^{-z}g\left( x\right) \,,
\nonumber
\\&
C({\bf k},\omega)=k^{-(2 \chi+d+z)}n\left( x\right) \,,
\nonumber
\end{eqnarray}
\noindent
with $x=\omega/k^z$. 
In $d=1$, the mode-coupling equations simplify, due to the 
existence of a fluctuation dissipation theorem which relates 
the correlation function to the response function. In $t$ and $k$
space, the fluctuation dissipation theorem can be written as: 
\begin{equation}
G(k,t)=\frac{\nu k^2}{D} \Theta(t) C(k,t) \,.
\end{equation}
(We use the same notation $G$ and $C$ in $t$ space and $\omega$ space and 
indicate which one we mean by the arguments).
We choose $\nu=1$ and $D=1$ in what follows. 
The mode-coupling equations are then reduced to one single equation 
that in terms of the response function in $k$ and $t$ space 
$G(t k^z)=G(k,t)$ reads:
\begin{equation}
G(\tau)=1-\lambda^2\int^{\tau}_{0}d\sigma \int_0^{\sigma}ds \nu(s)
G(\sigma-s) \,,
\end{equation}
where $\tau$ is the scaling variable $t k^z$, 
\begin{equation}
\nu(s)=\frac{1}{2 \pi}\int_{0}^{\infty}dx G(\mid 1/2+x \mid^z s)
G(\mid 1/2-x \mid^z s) \,,
\end{equation}
and $z=3/2$.
A similar analysis to the one we have done can be found in \cite{FTH}, 
where similar results were found (compare Fig. 1 with Fig. 1 in \cite{FTH}), 
but here we focus on the long time asymptotics.
To make this comparison we set here and in what follows $\lambda=1$.  
In a previous study \cite{uspre}, we argued that 
an asymptotic solution $\widehat{n}(\tau)=
\widehat{n_\infty}(\tau)$ for $\tau\rightarrow \infty$ for the 
scaling part of the correlation function in $t$ and $\bf{k}$ 
space ($\tau=t k^z$, and $\hat{n}(\tau)$ is the Fourier transform 
of $n(x)$) is given by 
\begin{equation}
\widehat{n_\infty}(\tau)=
A \,(B \tau)^{\frac{\gamma}{z}}
e^{-\!\mid B \tau \mid^{\frac{\alpha}{z}}}\,,
\label{largep}
\end{equation}
with $\gamma=(d-1)/2$, $\alpha=1$, and 
\begin{equation}
A=\frac{g(0)^{-2}4 \Gamma(4z-4)}{P 2^{(d-1)/2}
\Gamma(2z-2)^2}\,, 
\label{A}
\end{equation}
with
$P\!=\!\lambda^2/(2^d\pi^{(d+3)/2})$.
\begin{figure}
\narrowtext\centerline{\epsfxsize\columnwidth\epsfbox{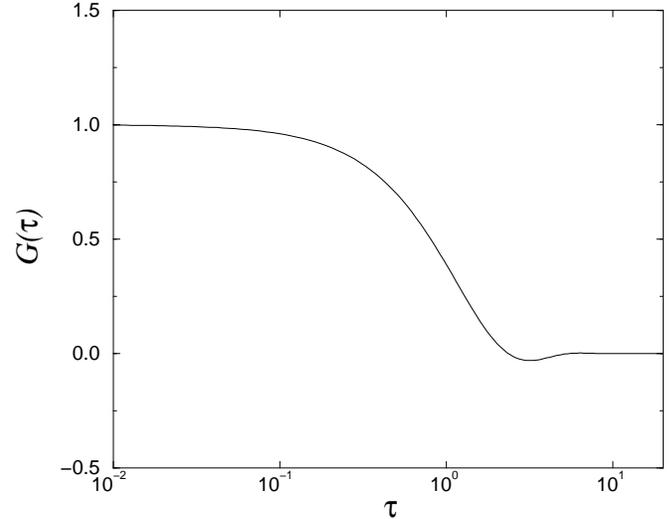}}
\caption{Scaling function for the response function $G(\tau)$.}
\end{figure}
In $d=1$, $G(\tau) \propto \hat{n}(\tau)$ due to the 
fluctuation dissipation theorem, so that we expect the same 
asymptotic expression for $G$. 
Our numerical analysis shows, much to our surprise, an  oscillatory behavior 
for the correlation function superimposed on the stretched exponential decay, 
a feature that has never been observed before. 
\begin{figure}
\narrowtext\centerline{\epsfxsize\columnwidth\epsfbox{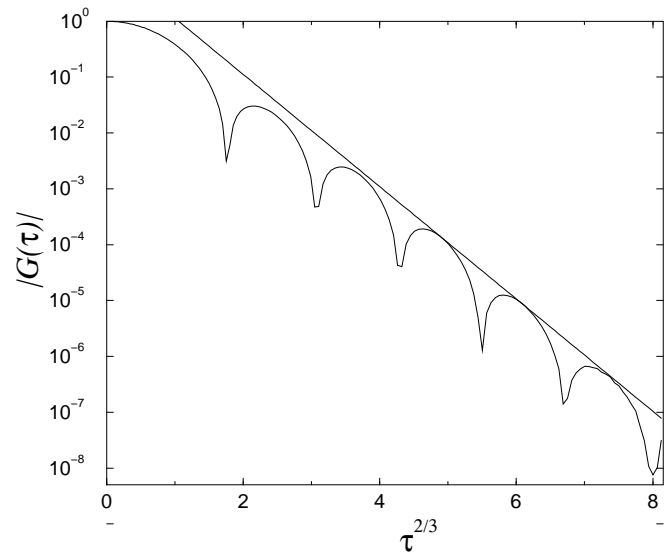}}
\caption{Oscillation and stretched exponential behavior in 
the response function $G(\tau)$. The dashed line indicates a line of slope 
$-1$ as predicted by Eq. (7). }
\end{figure} 
The long time behaviour can be revealed by plotting $\mid \!\!G(\tau)\!\!\mid$
as a function of $\tau^{2/3}$ in a linear-log scale (see Fig. 2), where both 
the presence of the oscillations and the overall stretched exponential 
decay of the envelope become apparent. 
We do not have any simple argument to explain 
the presence of such oscillations. However, we can show that
they are perfectly 
consistent with the saddle point analysis performed in \cite{uspre}. 
The same calculation can in fact be repeated with a complex exponential
with the constant $B$ a complex number. 
While the calculation leads to the same values of $\gamma$ and $\alpha$, 
it is not now possible to predict the amplitude constant $A$. 

We next compare the results from mode-coupling with the result 
for the scaling function in $d=1$ from the exact solution, which
recently has became available \cite{PS}. The result in \cite{PS}
is given in terms of a function $f(w)$ which is related to our 
$G(\tau)$ 
by 
\begin{equation}
f(w)=\frac{1}{\pi}\int_0^{\infty} dy \cos(wy) G(y^z/4) \,.
\end{equation} 
The results are shown in Fig. 3, and show a reasonable agreement 
between the mode-coupling approximation and the exact solution \cite{PS2}. 
From the exact solution it is also possible to numerically 
calculate $G(\tau)$ and compare it
with our approximate mode-coupling solution. The
exact solution also displays the oscillatory behavior which we have
discovered in the mode-coupling approximation \cite{PS2}. 

Note that the function $f(w)$ ($g^{''}(w)/8$ in \cite{PS}) is also related 
to the truncated correlation function in real space 
\begin{equation}
{\tilde{C}}(x,t)\equiv\int_{-\infty}^{\infty}\frac{dk}{\pi}
(C(k,0)-\rm{cos}(kx)C(k,t))=x F(t/x^{3/2})
\end{equation}
by $f(w)= \frac{1}{2}\frac{d^2}{dw^2}\left[w F(1/4 w^{2/3})\right]$. 

An earlier study of the accuracy of the mode-coupling approximation was
undertaken by Frey et al.\cite{FTH}, who studied the magnitude of the
corrections of higher order diagrams to the bare vertex. They found that
such diagrams  did produce substantial corrections. It is clear however
that such contributions are relatively unimportant for the correlation
function we have studied. 

In summary, we have presented a numerical study 
of the mode-coupling equations
for the strong coupling regime of the KPZ equation
in the long time limit in $d=1$. We recovered the stretched 
exponential relaxation
for the correlation function predicted previously in \cite{uspre}, 
but found a superimposed oscillation. Such oscillations are 
consistent with our previous analysis, even though we had not 
anticipated them. Finally, we compared 
the results from mode-coupling theory in $d=1$ with 
the exact solution from Ref. \cite{PS}.  

\begin{figure}
\narrowtext\centerline{\epsfxsize\columnwidth\epsfbox{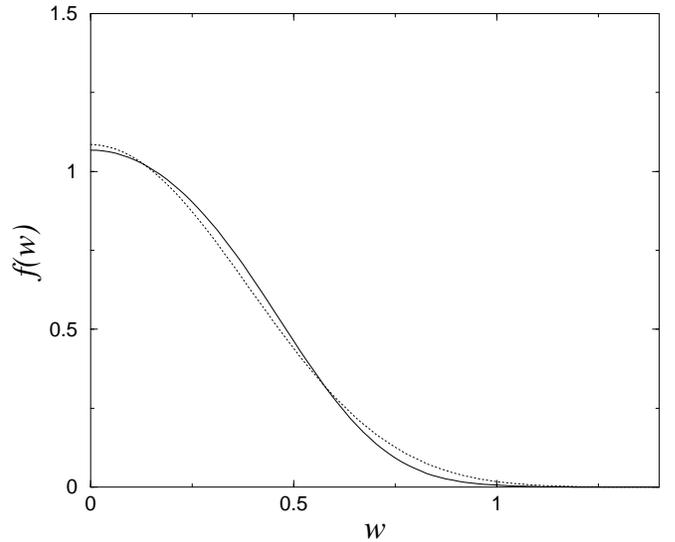}}
\caption{Comparison with the results in [10]: the solid line is 
our result for $f(w)$, the dotted line is the same function from Ref. [10].}
\end{figure}

The authors acknowledge  the support of EPSRC under grants 
GR/L38578 and GR/L97698. We thank M. Pr\"ahofer and H. Spohn
for sending us their data used in Fig. 3, and for discussions. 

\end{multicols}
\end{document}